\documentclass[preprint,superscriptaddress,citeautoscript,floatfix]{revtex4-1}
\usepackage{amssymb}
\usepackage{amsbsy}
\usepackage{amsmath}
\usepackage{bm}
\usepackage{color}
\usepackage{textcomp}
\usepackage{gensymb}
\usepackage[breaklinks,colorlinks = true,linkcolor = blue,urlcolor=cyan,citecolor=blue]{hyperref}
\usepackage{multirow}
\usepackage{array}
\usepackage{booktabs}
\usepackage{upgreek}
\usepackage{epsfig,psfrag,amsopn}
\usepackage{mathrsfs}
\usepackage{float}
\usepackage{graphicx}
\usepackage{subcaption}
\usepackage{tabularx}
\usepackage{subcaption}

\begin{document}

\title{Raman fingerprints of fractionalized Majorana excitations in honeycomb iridate Ag$_3$LiIr$_2$O$_6$}

\author{Srishti Pal}
\affiliation{Department of Physics, Indian Institute of Science, Bengaluru 560012, India}

\author{Vinod Kumar}
\affiliation{Department of Physics, Indian Institute of Technology Bombay, Mumbai 400076, India}

\author{Debendra Prasad Panda}
\affiliation{Chemistry and Physics of Materials Unit, Jawaharlal Nehru Center for Advanced Scientific Research, Jakkur P. O., Bengaluru 560064, India}

\author{A. Sundaresan}
\affiliation{Chemistry and Physics of Materials Unit, Jawaharlal Nehru Center for Advanced Scientific Research, Jakkur P. O., Bengaluru 560064, India}

\author{Avinash V. Mahajan}
\affiliation{Department of Physics, Indian Institute of Technology Bombay, Mumbai 400076, India}

\author{D. V. S. Muthu}
\affiliation{Department of Physics, Indian Institute of Science, Bengaluru 560012, India}

\author{A. K. Sood}
\email[E-mail:~]{asood@iisc.ac.in}
\affiliation{Department of Physics, Indian Institute of Science, Bengaluru 560012, India}

\date{\today}

\begin{abstract}

{We report low-temperature (down to $\sim$5 K) Raman signatures of the recently discovered intercalated honeycomb magnet Ag$_3$LiIr$_2$O$_6$, a putative Kitaev quantum spin liquid (QSL) candidate. The Kitaev QSL is predicted to host Majorana fermions as its emergent elementary excitations through a thermal fractionalization of entangled spins $S = 1/2$. We observe evidence of this fractionalization in the low-energy magnetic continuum whose temperature evolution harbours signatures of the predicted Fermi statistics obeyed by the itinerant Majorana quasiparticles. The magnetic Raman susceptibility evinces a crossover from the conventional to a Kitaev paramagnetic state below the temperature of $\sim$80 K. Additionally, the development of the Fano asymmetry in the low frequency phonon mode and the enhancement of integrated Raman susceptibilities below the crossover temperature signifies prominent coupling between the vibrational and Majorana fermionic excitations.} 

\end{abstract}

\maketitle

\section{Introduction}

The advent of quantum spin liquids (QSL) has demonstrated the potential of condensed matter systems to host elusive low energy excitations like Majorana fermions~\cite{Kasahara2018,Alicea2012}. The Kitaev QSL with spins $S = 1/2$ on two-dimensional (2D) honeycomb lattice offers an ideal platform in this context due to the thermal fractionalization of the highly entangled spin excitations into pairs of non-interacting itinerant Majorana quasiparticles coupled to localized $Z_2$ gauge fluxes~\cite{Kitaev2006,Knolle2014}. These two kinds of Majorana excitations, itinerant and localized, have well separated crossover
temperature scales, the higher $T_h \sim$0.4 -- 0.6 $J_K$ ($J_K$ is the Kitaev coupling strength) related to formation of matter Majorana fermions and the lower $T_l \sim$0.012 -- 0.015 $J_K$ associated with condensation of localized flux~\cite{Nasu2015,Nasu2016}.

Realization of Kitaev essence in real systems, as proposed in the pioneering works by Jackeli and Khaliullin~\cite{Jackeli2009} requires an intricate interplay of electronic (Coulomb) correlations, crystal-field effects, and spin-orbit entanglement as seen in certain heavy 4$d$ and 5$d$ transition metal compounds. Among these, the most noticeable ones are Na$_2$IrO$_3$~\cite{Mehlawat2017,Chun2015}, $\alpha$-RuCl$_3$~\cite{Do2017,Sandilands2015,Glamazda2017}, $\alpha$, $\beta$, and $\gamma$-Li$_2$IrO$_3$~\cite{Glamazda2016,Li2020}. Despite manifesting experimental signatures of fractionalization and hence showing proximity to Kitaev QSL state, all of these materials eventually exhibit long-range magnetic ordering at low temperatures~\cite{Winter2017}. This raises concern regarding the crucial role of non-Kitaev terms in the interaction Hamiltonian of the Kitaev systems affecting their pristine low-temperature QSL state.

Recently, a new set of Kitaev materials like H$_3$LiIr$_2$O$_6$~\cite{Kitagawa2018}, Cu$_2$IrO$_3$~\cite{Abramchuk2017}, and Ag$_3$LiIr$_2$O$_6$ have been synthesized from the parent compounds A$_2$IrO$_3$ (A = Na, Li) with the caveat that these compounds are susceptible to quenched disorders in their structure. In fact, the nature of disorder in these candidates becomes a key precursor in controlling the fate of their QSL ground state~\cite{Pei2020,Kenney2019,Pal2020,Bahrami2021}.

In this report, we study the Raman scattering signatures of the recently synthesized~\cite{Bette2019} Kitaev QSL candidate Ag$_3$LiIr$_2$O$_6$. Derived from its precursor $\alpha$-Li$_2$IrO$_3$ by replacing the inter-layer Li atoms with silver retaining the honeycomb network of the LiIr$_2$O$_6$ ($ab$) planes, Ag$_3$LiIr$_2$O$_6$ crystallizes in base centred monoclinic $C2/m$ structure. The LiO$_6$ octahedral connectivity linking the successive honeycomb planes in $\alpha$-Li$_2$IrO$_3$ are replaced by perfectly linear (180$^{\circ}$) O-Ag-O dumbbell bonds in Ag$_3$LiIr$_2$O$_6$, resulting in an $\sim$30\% increase in the interlayer separation. The initial magnetic and thermodynamic studies on disordered Ag$_3$LiIr$_2$O$_6$ samples with extended Ag positional defects within the honeycomb layers, entrenched the validity of Kitaev QSL physics in this system by affirming the absence of any long-range magnetic order along with a two-step release of spin entropy at crossover temperatures $T_h \sim$ 75 K and $T_l \sim$13 K~\cite{Bahrami2019}. Later cleaner batch of samples exhibited a more pronounced and clear transition to long-range ordering below the Neel temperature $T_N \sim$ 10 K while spin fluctuations remain present even in the ordered state down to 4.2 K~\cite{Bahrami2021,Wang2021,Mahajan2021}.

Our measurements unveil the essence of fractionalization in Ag$_3$LiIr$_2$O$_6$ typified by the broad low energy Magnetic Raman continuum attributable to deconfined Majorana quasiparticles obeying the predicted Fermi statistics. The presence of Majorana fermionic excitations in the system is further evidenced by their coupling with the low-frequency phonon mode causing development of clear Fano asymmetry in the line shape below the Majorana crossover temperature. Additionally, the Majorana-phonon coupling in the Kitaev paramagnetic phase is manifested through the departure of Raman intensity from the conventional thermal Bose contribution.

\section{Material Characterization and Experimental Methods}

High-quality polycrystalline samples of Ag$_3$LiIr$_2$O$_6$ were synthesized with the ion exchange method in two steps. In the first step the starting compound Li$_2$IrO$_3$ was prepared with conventional solid state reaction method. The stoichiometric amounts of initial chemicals Li$_2$CO$_3$ (Alfa Aesar, 99.998\%) and Ir-metal powder (Arro Biochem, 99.99\%) was taken with 5\% extra Li$_2$CO$_3$. The mixture was grounded well and heated at 650$^{\circ}$C for 15 Hrs in open air with the heating rate of 40$^{\circ}$C per hour. The mixture was again heated, with thorough grinding, at 950$^{\circ}$C and 1000$^{\circ}$C for 24 hours each. In the next step, Li$_2$IrO$_3$ was mixed with AgNO$_3$ (Alpha Aesar, 99.998 \%) in a molar ratio of 1:3 and after mixing for 10 minutes this mixture was heated at 200$^{\circ}$C for 12 hours with natural cooling. The final compound Ag$_3$LiIr$_2$O$_6$ was obtained after washing it multiple times with deionized water to remove the excess AgNO$_3$ and reaction product LiNO$_3$. The presence of nitrates was checked with KCl solution.

\begin{figure}[h!]
\centering
\includegraphics[width=160mm,clip]{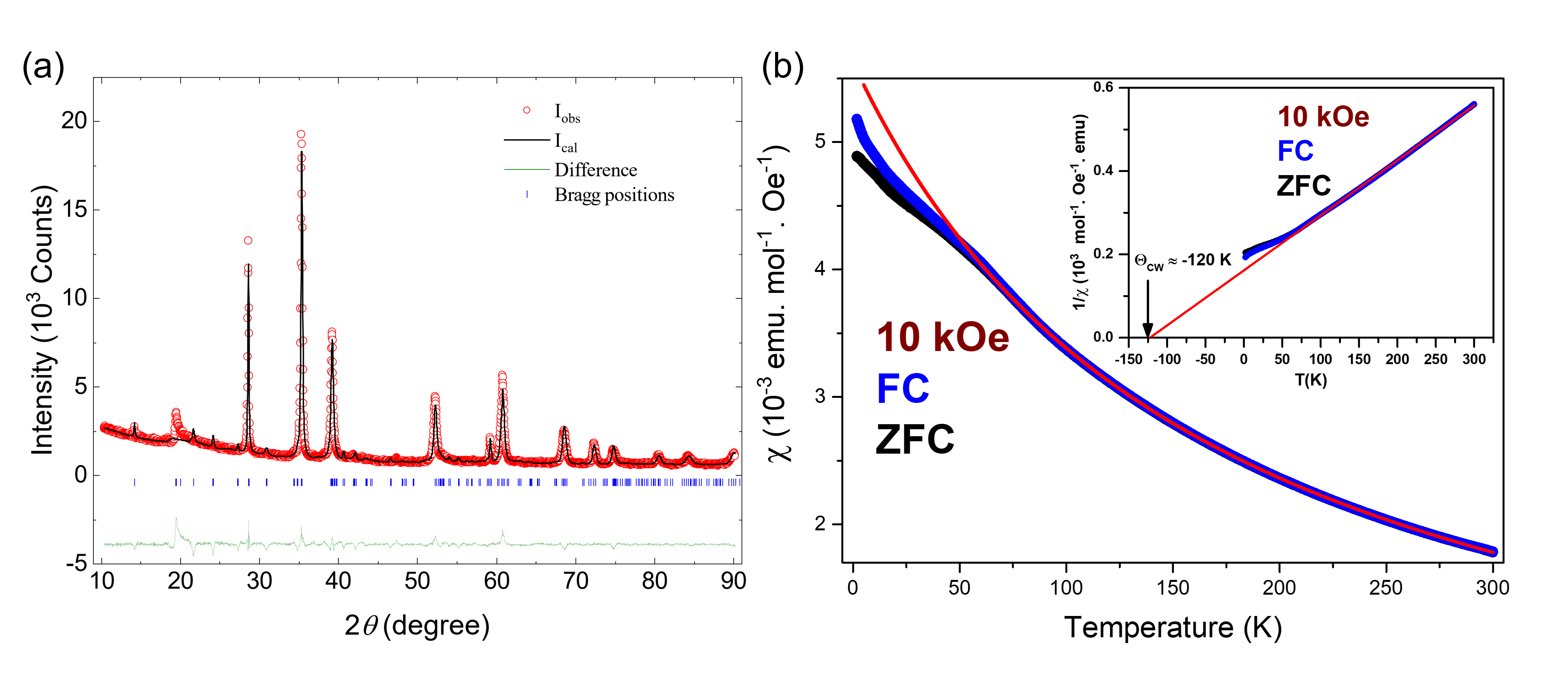}
\caption{\small (a) Rietveld refined powder x-ray diffraction pattern of Ag$_3$LiIr$_2$O$_6$. Experimental data and calculated pattern are shown by red circles and black solid line, respectively. Reflection positions for are indicated by blue vertical bars and the weighted difference between observed and calculated profiles is shown by the lower dark green curve. (b) Temperature dependence of the \textit{dc} magnetic susceptibility between $2$ and $300$~K at a field value of 1 T. Field cooled (FC) and zero field cooled (ZFC) data are shown by blue and black circles, respectively. The red solid line is the Curie-Weiss (CW) fit. Inset shows the temperature variation of inverse susceptibility with the Curie-Weiss temperature extracted as  120 K.}
\label{figS1}
\end{figure}

The powder X-ray diffraction measurements were performed with a PANalytical Empyrean diffractometer having a Cu target with $K_{\alpha}$ radiation ($\lambda$ = 1.54182 \AA). The Rietveld refinement done using \textsc{fullprof} suite shown in Fig.~\ref{figS1}(a) confirms phase purity with the monoclinic \emph{C2/m} crystallographic structure. We performed refinement without considering stacking fault by excluding the region 18-21$^{\circ}$ (the Warren peak position). The extracted cell parameters and the atomic positions are given in Table~\ref{tableS1}.

\begingroup
\squeezetable
\begin{table*}[htb!]
%\addtolength{\tabcolsep}{-1.5pt}
\caption{\small Rietveld refined parameters.}
\centering
 \begin{tabularx}{\textwidth}{c c c c c c c c} 
 \hline \hline \\ [-1ex]
 \textbf{Space group} & \textbf{Lattice parameters} & \textbf{Refinement parameters} & \textbf{Atom} & $\boldsymbol{x}$ & $\boldsymbol{y}$ & $\boldsymbol{z}$ & \textbf{Occupancy}\\[0.5ex]
\hline \\ [-1ex]
  \multirow{6}{1cm}{\centering{$C2/m$}} & $a$ = 5.2754(2) \AA & {} & Ir1 & 0.0000 & 0.3341(5) & 0.0000 & 1 \\[0.5ex]
  {} & $b$ = 9.1370(6) \AA & $R_p$ = 20.3\% & Li1 & 0.0000 & 0.0000 & 0.0000 & 1 \\[0.5ex]
  {} & $c$ = 6.4827(4) \AA & $R_{wp}$ = 17.6\% & Ag1 & 0.0000 & 0.1725(7) & 0.5 & 1 \\[0.5ex]
  {} & $\alpha$ = 90$^{\circ}$ & $R_{exp}$ = 7.03\% & Ag2 & 0.0000 & 0.5 & 0.5 & 1 \\[0.5ex]
  {} & $\beta$ = 105.73(1)$^{\circ}$ & $\chi^2$ = 6.25 & O1 & 0.0269(1) & 0.0000 & 0.0198(1) & 1 \\[0.5ex]
  {} & $\gamma$ = 90$^{\circ}$ & {} & O2 & 0.4199(51) & 0.3429(36) & 0.1673(26) & 1 \\[0.5ex]
\hline \hline
\end{tabularx}
\label{tableS1}
\end{table*}
\endgroup

The magnetic susceptibility measurements were performed using Magnetic Property Measurement System (MPMS-SQUID, Quantum Design, USA) in Vibrating Sample Magnetometer (VSM) mode under an applied field of 1 T. In zero-field-cooled (ZFC) measurement, the sample was first cooled to lowest temperature (2 K) without applying magnetic field and the measurements were carried out in warming cycle. In field-cooled (FC) mode, sample was cooled under the applied magnetic field and the magnetization was measured on warming. Figure~\ref{figS1}(b) shows the temperature variation of the \textit{dc} magnetic susceptibility between $2$ and $300$~K. The high temperature $\chi$ ($75 < T < 300~K$) is fitted well with the modified Curie-Weiss (CW) form $\chi = \chi_0 + C/(T-\Theta_{CW})$. The fit deviates below $\sim$50 K along with a bifurcation between the ZFC and FC curves as reported earlier~\cite{Mahajan2021} to be originated from the quenched disorder of naturally occurring stacking faults present in the sample inducing localized moments. Inset of Fig.~\ref{figS1}(b) shows the inverse \textit{dc} susceptibility giving the Curie-Weiss temperature $\Theta_{CW} \approx -120$ K.

Unpolarised micro-Raman measurements were performed in back-scattering geometry using Horriba LabRAM HR Evolution Spectrometer equipped with thermoelectric cooled charge coupled device (CCD) detector (HORIBA Jobin Yvon, SYNCERITY 1024 $\times$ 256). The sample was excited using 532 nm DPSS laser with $\sim$190 $\mu$W power falling on the sample. Temperature variation from 5 K to 293 K with a temperature stability of $\approx \pm$1 K was done with closed cycle He cryostat (Cryostation S50, Montana Instruments).

\section{Results and Discussion}

\begin{figure}%[h!]
\centering
\includegraphics[width=80mm,clip]{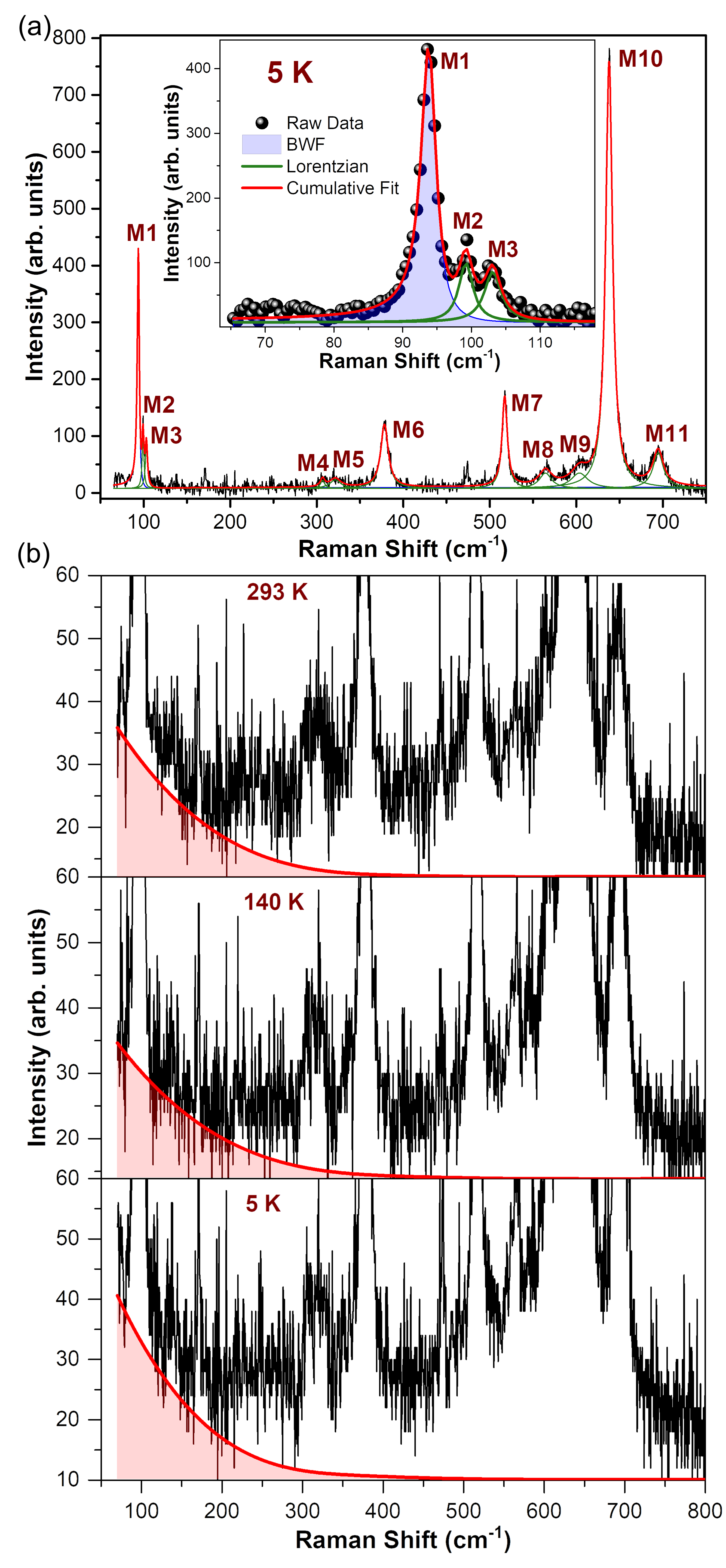}
\caption{\small (a) Raman spectra of Ag$_3$LiIr$_2$O$_6$ at 5 K in the spectral range 70 - 750 cm$^{-1}$. Black, blue, green, and red curves denote the raw data, phonon fits with Breit-Wigner-Fano (BWF) and Lorentzian line shapes, and the cumulative fit, respectively. Inset shows the asymmetry in the line shape of M1 mode at 5 K fitted with the BWF function. (b) Magnified Raman profile at selected temperatures with the magnetic continuum shown by the shaded regions.}
\label{fig1}
\end{figure}

Ag$_3$LiIr$_2$O$_6$ crystallizes in monoclinic space group $C2/m$ (\# 12) with 2 formula units ($Z$ = 2) per unit cell. Factor group analysis gives 15 Raman active $\Gamma$-point phonon modes as, $\Gamma_{Raman} = 7A_g + 8B_g$ among which 11 modes could be detected at 5 K in our unpolarized Raman experiments denoted by M1-M11 in Fig.~\ref{fig1}(a). The lowest frequency M1 mode unveils clear asymmetry in its line shape at low temperatures and hence is fitted with the asymmetric Breit-Wigner-Fano (BWF) profile~\cite{Hasdeo2014} as shown in the inset of Fig.~\ref{fig1}(a). The BWF function is defined as,
\begin{equation}
I_{BWF}(\omega_s) = I_0\frac{[1+s/q]^2}{1+s^2}
\label{BWF}
\end{equation}
where $s=(\omega_s-\omega_0)/w$. The parameters $\omega_s$, $\omega_0$, $w$, $1/q$, and $I_0$ are the Raman shift, the spectral peak centre, the spectral width, the asymmetry factor, and the maximum intensity of the BWF line, respectively. The low-frequency region of the Raman profile is magnified and shown in Fig.~\ref{fig1}(b) at a few selected temperatures. The phonon modes are superimposed on a weak low energy continuum reminiscent of putative spin liquid materials~\cite{Sandilands2015,Glamazda2016,Glamazda2017,Li2020,Pal2020} as being originated from the scattering of light by exotic Majorana fermionic excitations which are fractionalized in nature. Like other derived QSL candidate Cu$_2$IrO$_3$~\cite{Pal2020}, Ag$_3$LiIr$_2$O$_6$ also lacks the peaking feature and the low energy linear omega dependence of Raman intensity at low temperatures as was seen in the first-generation candidates~\cite{Sandilands2015,Glamazda2016,Li2020} and hence masks the pristine low-frequency signatures of the Majorana fermions. This is owing to the intricate role of disorder present in these derived candidates which has been theoretically predicted to enhance the low energy density of states of the Majorana fermionic excitations~\cite{Willans2010,Willans2011}. Equating the upper cut off of $\sim$300 cm$^{-1}$ of this broad Raman response to 3$|J_K|$ ($J_K$ is the Kitaev interaction strength), the theoretical estimate for the band edge of the Raman continuum from Majorana fermions~\cite{Nasu2016}, yields $|J_K| \approx$ 12 meV which is in reasonable agreement with the Kitaev interaction strength obtained for this compound from recent density functional theoretical (DFT) calculations ($J_K \approx$ 11.4 meV)~\cite{Mahajan2021}.

\begin{figure}%[h!]
\centering
\includegraphics[width=160mm,clip]{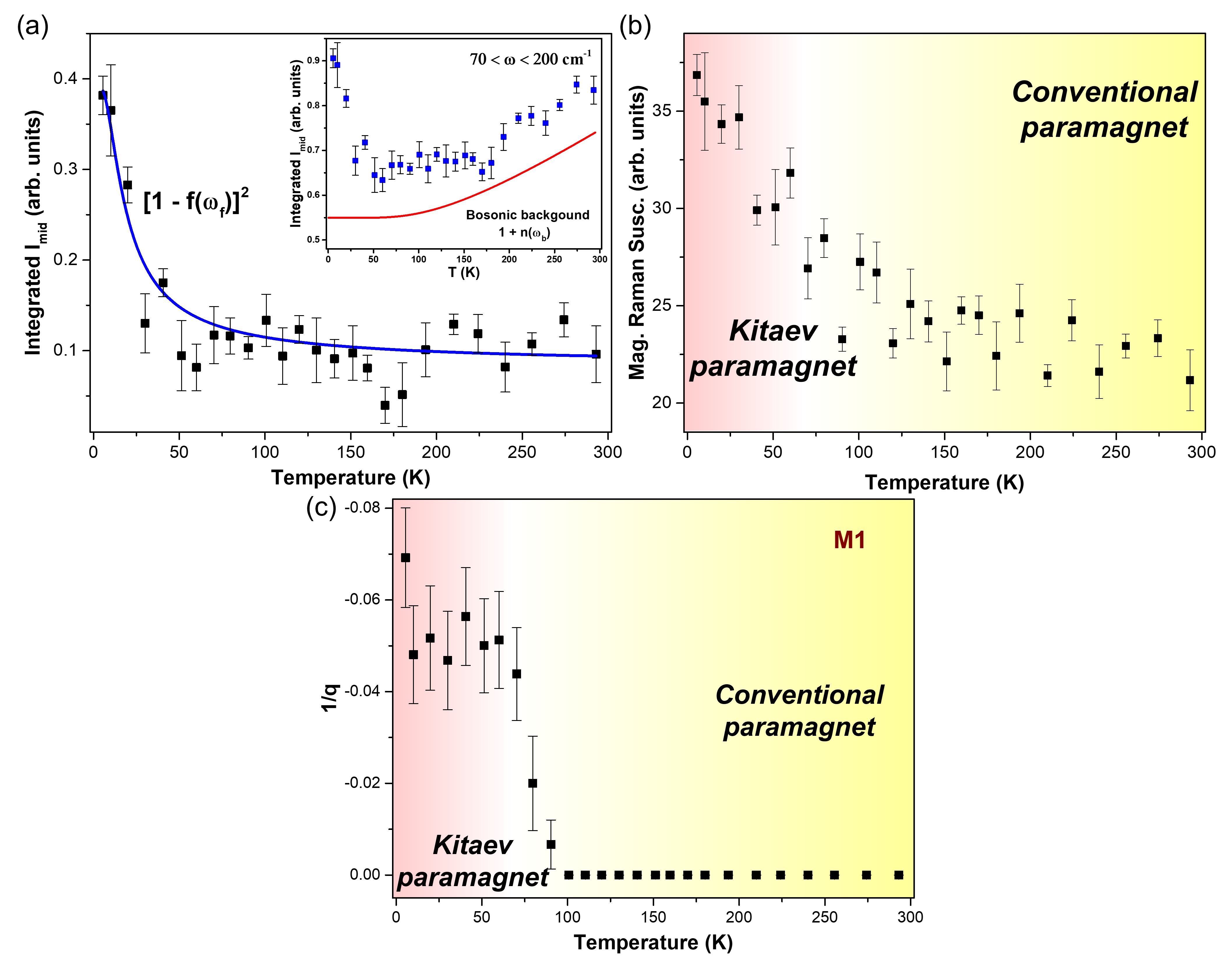}
\caption{\small (a) Integrated $I_{mid}$ as a function of temperature in the frequency range 70-200 cm$^{-1}$ after subtracting the bosonic background (shown in inset). The blue solid curve denotes fitting by the two-fermion scattering function $A+B(1 - f)^2$ [$f = 1/(1 + e^{\hbar \omega/k_BT})$ is the Fermi distribution function]. (b) Temperature evolution of magnetic Raman susceptibility calculates using the Kramers-Kronig relation. The shading marks the crossover between the conventional and Kitaev paramagnetic phases. (c) Temperature dependence of the asymmetry parameter $1/q$ of M1 mode. The shading indicates the boundary between the conventional and Kitaev paramagnetic states.}
\label{fig2}
\end{figure}

The smoking gun evidence for the emergence of the Kitaev QSL phase, as witnessed by all other Kitaev QSL materials~\cite{Nasu2016,Glamazda2016,Pei2020,Pal2020}, is the robust $(1-f)^2$ scaling (where $f(\omega) = 1/(1+e^{\hbar\omega/k_BT})$ is the Fermi distribution function) of the integrated Raman continuum in the mid frequency regime stemming from the creation of two itinerant Majorana fermions due to interaction with the probe photons. Figure~\ref{fig2}(a) shows the temperature evolution of $I_{mid}$ integrated in the frequency interval of 70 to 200 cm$^{-1}$ chosen in accordance with the theoretical estimate given by Nasu et al.~\cite{Nasu2016} (see Appendix~\ref{A_2} for robustness of $I_{mid}$ frequency window). The red curve in the inset denotes the Bosonic background due to one particle scattering of the form $[1+n(\omega_b)] = 1/(1-e^{-\hbar \omega_b/k_BT})$, with $\omega_b = 35$ meV, dominating the high-temperature regime. The main panel of Fig.~\ref{fig2}(a) shows the temperature dependence of the magnetic contribution to integrated $I_{mid}$ after subtracting the thermal Bose contribution. The blue curve represents the fitting of the magnetic continuum to the two-fermion scattering form $A+B[1-f(\omega_f)]^2$, with $\omega_f = 2$ meV and reinforces the possibility of fractionalized Majorana fermionic excitations, corroborating Ag$_3$LiIr$_2$O$_6$ to be a potential spin liquid candidate.

Further exploration of the Majorana continuum requires extraction of the magnetic Raman susceptibility $\chi_R^{dyn}$, a measure of the dynamic response from the spin system. Figure~\ref{fig2}(b) exhibits the temperature dependence of $\chi_R^{dyn}$ deduced by integrating the Raman conductivity $\frac{\chi''(\omega)}{\omega}$ from 70 to 400 cm$^{-1}$ in accordance with the Kramers-Kronig relation, $\chi_R^{dyn} = \lim_{\omega \to 0} \chi(k=0, \omega)\equiv \frac{2}{\pi}\int{\frac{\chi''(\omega)}{\omega}d\omega}$. According to the fluctuation-dissipation theorem, the dynamical Raman tensor susceptibility $\chi''(\omega)$ is related to the Raman intensity $I(\omega)$ as, $I(\omega) = 2\pi \int{\left\langle R(t)R(0) \right\rangle e^{i\omega t}dt} \propto [1+n(\omega)]\chi''(\omega)$, where $R(t)$ is the Raman operator which within the Kitaev QSL phenomenology, directly couples to the dispersing fractionalized Majorana fermions and hence, substantially projects the density of states (DOS) of weighted two-Majorana fermions. As seen from Fig.~\ref{fig2}(b), $\chi_R^{dyn}$ of Ag$_3$LiIr$_2$O$_6$ remains almost constant down to $\sim$80 K below which it increases rapidly signifying moderate enhancement of the Majorana fermionic DOS driving the system to a Kitaev paramagnetic phase as also seen in other Kitaev QSL candidates $\alpha$-RuCl$_3$~\cite{Glamazda2017} and $\alpha$-Li$_2$IrO$_3$~\cite{Li2020}. Also, the Majorana crossover temperature $T_h \: \sim$80 K $\approx \: 0.6 \; J_K$ as gleaned from the temperature evolution of $\chi_R^{dyn}$ matches well with the theoretical predictions by Nasu et al.~\cite{Nasu2015,Nasu2016}. This value of $T_h$ is in agreement with the earlier reported $T_h$ of $\sim$75 K by Bahrami et al.~\cite{Bahrami2019,Bahrami2021} from heat capacity measurements on Ag$_3$LiIr$_2$O$_6$ system.

The temperature evolution of the $1/q$ asymmetry parameter for the M1 mode is shown in Fig.~\ref{fig2}(c). The asymmetry parameter remains essentially zero down to the crossover temperature of $T_h \sim$80 K below which it increases rapidly in magnitude with decreasing temperature. The emergence of this Fano resonance below $T_h$ has its root in the strong coupling between the discrete phonon mode (M1) and the magnetic continuum offered by the fractionalized Majorana excitations as also been seen in other Kitaev QSL candidates~\cite{Sandilands2015,Glamazda2017,Pal2020}. All other phonon modes (M2-M11) are fitted with the symmetric Lorentzian line shape for the entire range of temperature. The phonon spectra do not reveal emergence of new modes or disappearance of existing modes, confirming the absence of any structural transition throughout the entire range of temperature. 

\begin{figure}%[h!]
\centering
\includegraphics[width=160mm,clip]{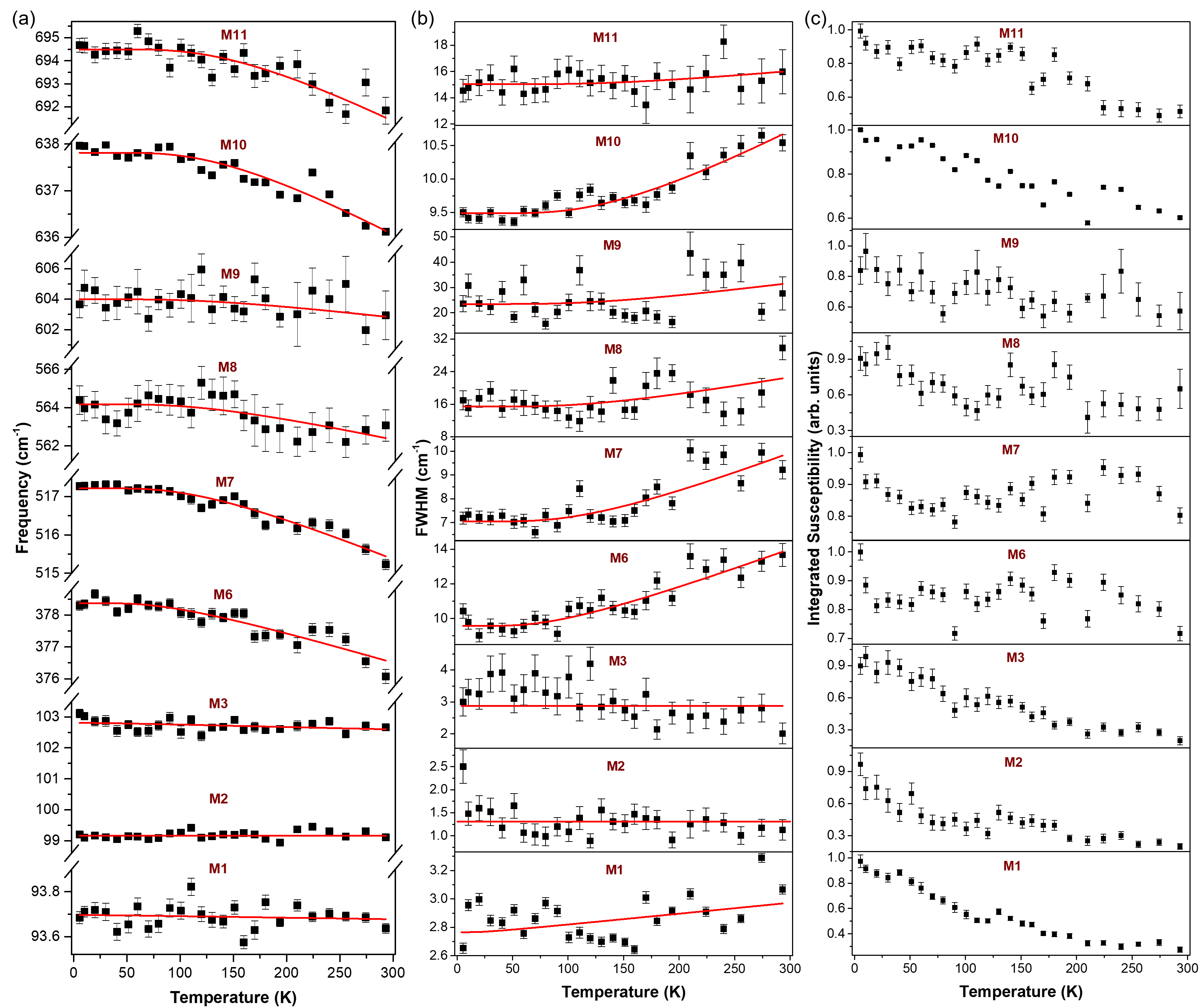}
\caption{\small Temperature evolution of (a) frequency, (b) FWHM, and (c) integrated susceptibilities of selected phonon modes. Red lines are the anharmonic fits to phonon frequencies and FWHMs. Blue lines are guide to eye indicating departures of the integrated susceptibilities below the Kitaev crossover temperature (shown by the shaded regions).}
\label{fig3}
\end{figure}

Figures~\ref{fig3}(a)-(c) showcase the temperature evolution of the frequencies, line widths, and the integrated susceptibilities of the strong phonon modes. The changes in frequencies and FWHMs, the real and imaginary parts of the phonon self-energy, of all the modes follow the typical monotonic lattice anharmonicity, as indicated by the simplified cubic anharmonic fits (red solid lines), arising from the decay of an optical phonon of frequency $\omega_0$ into a phonon pair ($\omega_0/2$). See Appendix~\ref{A_3} for more details. All the phonon modes including the asymmetric Fano mode (M1) exhibit no noticeable anomaly in their temperature dependence of frequencies and linewidths as compared to the sister compound Cu$_2$IrO$_3$~\cite{Pal2020}. However, the integrated susceptibilities of most of the phonon modes show anomalous rise with decreasing temperatures authenticating the development of spin-spin correlations in the Kitaev paramagnetic phase which affects the vibrational intensities through transfer of magnetic dipole intensity to the phonons due to significant coupling between the vibrational and fractionalised Majorana degrees of freedom~\cite{Allen}.

\section{Conclusions}

In summary, we have scrutinized the Raman scattering signatures of the proximate Kitaev QSL candidate Ag$_3$LiIr$_2$O$_6$. The low-energy broad magnetic continuum, following the expected Fermi statistics in its temperature evolution, offers a Kitaev coupling strength of $J_K \approx$ 12 meV, consistent with the value gleaned from the recent DFT calculations. The Majorana crossover temperature of $T_h \sim$80 K has been extracted from the development of dynamic spin susceptibility and is further supported by emergence of Fano asymmetric line shape for the lowest frequency phonon mode. Also, the intensity of most of the Raman phonons show prominent rise with decreasing temperatures. Our results thus provide the very first scattering signatures for the Kitaev spin liquid phase in the honeycomb iridate Ag$_3$LiIr$_2$O$_6$ and establish Ag$_3$LiIr$_2$O$_6$ as an ideal avenue to explore spin fractionalization and Majorana-phonon coupling.

\begin{acknowledgments}

{AKS thanks Nanomission Council and the Year of Science professorship of DST for financial support. AS acknowledges International Centre for Materials Science, Jawaharlal Nehru Centre for advanced Scientific Research, for providing experimental facility. VK and AVM acknowledge the central facilities provided by IIT Bombay and IRCC. AVM thanks MHRD for funding of STARS research project ID 358.}

\end{acknowledgments}

\appendix

\section{\bf{integrated $I_{mid}$}}
\label{A_2}

\begin{figure}[h!]
\centering
\includegraphics[width=100mm,clip]{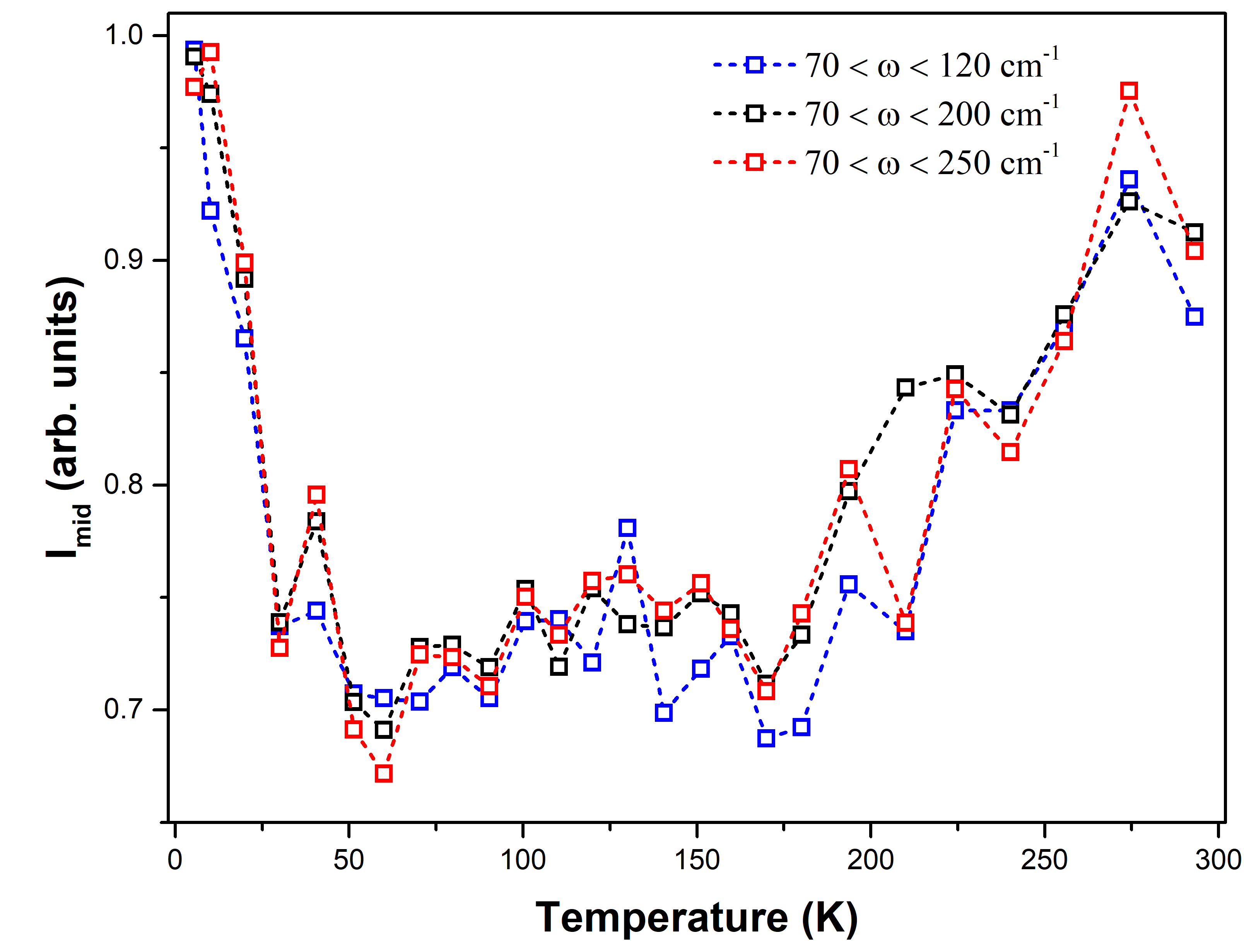}
\caption{\small Temperature dependence of integrated $I_{mid}$ for different energy ranges.}
\label{figS2}
\end{figure}

Figure~\ref{figS2} shows the temperature evolution of integrated $I_{mid}$ (normalized) in different energy windows. The typical scaling form of integrated $I_{mid}$ remains robust under these moderate variations of the energy range.

\section{Lattice anharmonicity}
\label{A_3}

{Temperature dependence of the phonon population is known as intrinsic anharmonic effect. Below the Debye temperature $\theta_D$ of the system, restricting to cubic (third-order) corrections to phonon self-energy, the simplest decay channel is offered by each phonon of frequency $\omega_0$ decaying into two phonons of equal energy \textit{i.e.}, $\omega_1 = \omega_2 = \frac{\omega_0}{2}$. Under this cubic anharmonicity, the phonon frequency $\omega (T)$ and FWHM $\Gamma (T)$ can be given as~\cite{Klemens1966},

\begin{equation}
\omega(T)= \omega_0 + A[1+2n(\frac{\omega_0}{2})]
\end{equation}

\begin{equation}
\Gamma(T)= \Gamma_0 + B[1+2n(\frac{\omega_0}{2})]
\end{equation}

where, $\omega_0$ and $\Gamma_0$ are constants at $T = 0$ K, $n(\frac{\omega_0}{2})$ is the Bose-Einstein thermal population factor, and A (negative) and B (positive) are constants. In the fits shown in Fig.~\ref{fig3}(a)-(b), $\omega_0$ is extracted from the frequency fits and those values of $\omega_0$ are used in the fitting of FWHMs. The fitting parameters $\omega_0$, $\Gamma_0$, A, and B for different modes are given in Table~\ref{tableS2} below.}

\begingroup
\squeezetable
\begin{table}[!htb]
%\addtolength{\tabcolsep}{-1.5pt}
\caption{\small List of fitting parameters for the cubic anharmonic fits to the phonon modes of Ag$_3$LiIr$_2$O$_6$}
\centering
    \begin{tabularx}{\columnwidth}{X X X X X}
    \hline \hline \\ [-1ex]
    \textbf{Mode} & \boldsymbol{$\omega_0 (cm^{-1})$}  & \textbf{A} \boldsymbol{$(cm^{-1})$} & \boldsymbol{$\Gamma_0 (cm^{-1})$}  & \textbf{B} \boldsymbol{$(cm^{-1})$} \\ [0.5ex]
    \hline \\ [-1ex]
    M1 & 93.7 & 0 & 2.7 & 0.03 \\ [0.5ex]
    M2 & 99.2 & 0 & 1.3 & 0 \\ [0.5ex]
    M3 & 102.8 & - 0.03 & 2.9 & 0 \\ [0.5ex]
    M6 & 379.7 & - 1.4 & 6.3 & 3.3 \\ [0.5ex]
    M7 & 519.5 & - 2.3 & 3.5 & 3.5 \\ [0.5ex]
    M8 & 566.8 & - 2.6 & 5 & 10.5 \\ [0.5ex]
    M9 & 605.9 & - 1.9 & 9.9 & 13.5 \\ [0.5ex]
    M10 & 641 & - 3.2 & 7.2 & 2.2 \\ [0.5ex]
    M11 & 701.3 & - 6.8 & 12.8 & 2.2 \\ [0.5ex]
    \hline \hline
    \end{tabularx}
    \label{tableS2}
\end{table}
\endgroup

\end{document}